# Space-Charge Limited Schottky Diodes with Wide-Bandgap Thin-film Oxide Heterojunctions

Jihun Lim[1,2,3*]

[1]School of Semiconductor and Chemical Engineering, Jeonbuk National University, Jeonju, 54896 Republic of Korea

[2]Department of Department of Semiconductor Science and Technology, Jeonbuk National University, Jeonju, 54896 Republic of Korea

[3]Department of Electrical and Computer Engineering, University of Michigan, MI 48109, USA

[*]Jihun Lim **email**: jihunlim@jbnu.ac.kr

## Abstract

Non-crystalline thin-film Schottky diodes are cost-effective but often exhibit unreliable electrical characteristics due to material imperfections. In this work, I present a Schottky diode structure utilizing in-situ grown $Ta_2O_5$ and $ZnO$ thin films deposited at room temperature. The low conduction band offset across the interface of the heterogeneous oxides facilitates efficient electron injection under forward bias. Capacitance-voltage characterization reveals a robust Schottky barrier at the $Au/Ta_2O_5$ interface without a significant barrier thinning effect, enabling high-voltage breakdown up to 65V and a high on/off ratio of $1\times10^8$. In demonstrations, the thin-film structure shows Schottky contact characteristics using even a relatively low work function metal of ITO, allowing the operation of transparent Schottky diodes. The diodes show additional potential for applications, including RF-to-DC conversion leveraging space charge capacitance at the $Ta_2O_5$/ZnO junction and rectifying resistive random access memory devices. This work highlights a promising approach for integrating low-cost, high-reliability Schottky diodes into back-end-of-line processes for wireless electronics and power devices.

## Introduction

Schottky diodes are fundamental rectifying components used in diverse applications, including automobile electronics,[1,2] battery chargers,[3] and telecommunications.[4] High blocking voltage operations and reliability often necessitate the use of expensive materials such as GaN,[5] SiC,[3] and $Ga_2O_3$[1,2,6], which have wide bandgaps and high crystallinity. However, these materials present challenges due to their high cost, limited scalability, and complex manufacturing processes, which can hinder their integration into advanced, cost-effective electronics. To address these issues, researchers have explored the use of thin-film materials, such as oxide semiconductors,[7,8] graphene,[9-11] and transition metal dichalcogenides (TMDs),[1,9,12]. TMDs are particularly attractive due to their potential to form defect-free metal-to-semiconductor junctions, which can achieve nearly ideal Schottky barriers and enable reliable rectification. Graphene, with its tunable work functions and van der Waals heterostructures, has improved the formation of Schottky junctions with Si or TMDs.[9,10,13] However, the challenges remain with graphene and TMDs, including difficulties in achieving high-quality large-scale production, precise control of film thicknesses, and compatibility with existing semiconductor processes, which raises questions about their commercialization prospects.

In contrast, oxide materials such as Indium–gallium–zinc–oxide (IGZO) and zinc oxide (ZnO) have demonstrated flexible Schottky diodes with gigahertz cutoff frequencies ($f_c$).[7,8]

Especially, due to their compatibility with existing manufacturing processes, the materials support conventional mass productions in display markets.[14] However, oxide thin-film devices often deviate from the Schottky-Mott rule due to junction defects and Fermi-level pinning effect, which can result in relatively high reverse currents and low breakdown voltages.[10, 15]

In this study, I introduce $Ta_2O_5$/ZnO heterostructure generated through in-situ rf-sputtering deposition at room temperature. A Schottky barrier is formed at the interface between Au and $Ta_2O_5$, which is not suffered from barrier thinning effects and allows for high blocking voltages up to 65V. Under forward bias, the low conduction band offset at the $Ta_2O_5$/ZnO junction enables electron injection from ZnO to $Ta_2O_5$. The voltage primarily drops across $Ta_2O_5$, where the drift mechanism governs charge transport to the Au electrode. At the $Ta_2O_5$/ZnO interface, electrons accumulate and diffuse to the Au, resulting in space-charge limited current (SCLC) characteristics. By replacing Au with ITO, an all-oxide Schottky diode is demonstrated on glass substrates, with an average transmittance of approximately 80 % in visible wavelengths. The addition of a thin 5-nm-thick $TaO_x$ layer between $Ta_2O_5$ and ZnO creates both trap-filled and trap-free SCLC regimes, resulting in a resistive random-access memory (ReRAM) function with high resistance window exceeding $10^4$. Thus, the Schottky diode structure offers potential applications in low-cost, high-yield Schottky diodes, transparent electronics, and advanced memories.

## Experimental Section

**Device Fabrication**

To prepare a thin-film $Ta_2O_5$/ZnO Schottky diode, a 10-nm-thick Ti adhesion layer and a 100-nm-thick Au layer are sequentially deposited by sputtering onto a cleaned glass substrate.

The mesa structure of $Ta_2O_5$/ZnO double layers is then patterned using standard lift-off processes, followed by in-situ rf-magnetron sputtering deposition of $Ta_2O_5$ and ZnO targets (Kult J. Lesker) of 99.999% purity. A 100-nm ITO top-electrode is deposited by sputtering and patterned using liftoff processes. For an all-oxide transparent Schottky diode, the same processes are used, but the bottom electrode is replaced with ITO deposited by dc sputtering. To prepare a crossbar Schottky diode, a 10-nm-thick Ir and a 50-nm-thick Au layers are sequentially deposited by sputtering onto a cleaned glass substrate and patterned using standard liftoff processes. Four layers – $Ta_2O_5$ (200nm), $TaO_x$ (5nm), ZnO (50nm), and ITO (50nm) – are then deposited by in-situ sputtering and patterned using standard liftoff processes. The $TaO_x$ is deposited by reactive dc magnetron sputtering in a mixed atmosphere of Ar (10 sccm) and $O_2$ (30 sccm) pressures.

**Measurements**

The current-voltage and capacitance-voltage characteristics were measured using a Keithley 4200 semiconductor analyzer with source-measure units and capacitance-measure units. For half-wave rectification measurements, a waveform generator (Agilent 33220A) provided sine-wave input voltage to the diode, and the output voltage is recorded with an oscilloscope (Agilent DSO7054A). The transmittance of all-oxide Schottky diodes is measured using a spectroscopic ellipsometer (J.A. Woollam M–2000).

## Results and Discussion

**Characterization of $Ta_2O_5$/ZnO Hetero-oxide Schottky Diode**

Figure 1a illustrates the energy band structure of the $Ta_2O_5$/ZnO heterojunction. The Schottky barrier ($\phi_b$) is formed at the metal/$Ta_2O_5$ interface. According to the Schottky-Mott rule, ideal $\phi_b$ can range from 1.2 eV to 0.7 eV, depending on the work functions of metals. Figure 1b indicates the measurement using the Schottky diode. The ITO is grounded while the voltage is applied to the Au. The symbols represent the measured data using the diode, with the red solid lines indicating the SCLC characteristics following the relationship $J \propto V^2$ according to the Mott-Guttley law [16], influenced by the band offset and charge accumulation. The inset shows an optical image of the diode with the Au and ITO contacts at the bottom and top, respectively. It retains an on/off ratio of higher than $10^6$.

Figure 1c shows the current-voltage characteristics of Au/Ta$_2$O$_5$/ITO and Au/ZnO/ITO devices. The Ta$_2$O$_5$ device has a Fermi level close to the conduction band,[17] and exhibits diode behavior with bi-Schottky barriers at the Au and ITO contacts. Due to the relatively high work function of Au compared to ITO, the turn-on voltage ($V_T$) at forward bias is smaller than at reverse bias. A typical Au/ZnO device with low defects can show Schottky diode behavior [18], but the experiment shows ohmic behavior due to high defects in the ZnO layer.[19] Figure S1 shows the photoluminescence measurements of two different ZnO thin films on glass substrates, prepared by sputtering at room temperature and atomic layer deposition at 150 °C, respectively. By inserting the Ta$_2$O$_5$ between the Au and sputtering-ZnO layers, the ZnO acts as the electron transport layer to the Ta$_2$O$_5$ layer.

The reciprocal capacitance-voltage characteristics is investigated to estimate the built-in potential ($\phi_{bi}$) in the Schottky-diode structure. Figure 1d shows the equivalent circuit of the Schottky diode. The $C_f$ means free charges of electrons, $C_L$ is the localized trapped charges due to imperfect materials, and $R_L$ is the resistance responding to a given voltage by the trapped charges. From the imaginary impedance of the circuit (see the full deriviation in Note 1, Supporting Information), the frequency-dependent capacitance is expressed by:

$$C(\omega) = \frac{(C_L+C_f)^2+\omega^2 R_L^2 C_L^2 C_f^2}{(C_L+C_f)+\omega^2 R_L^2 C_L^2 C_f}, \qquad (1)$$

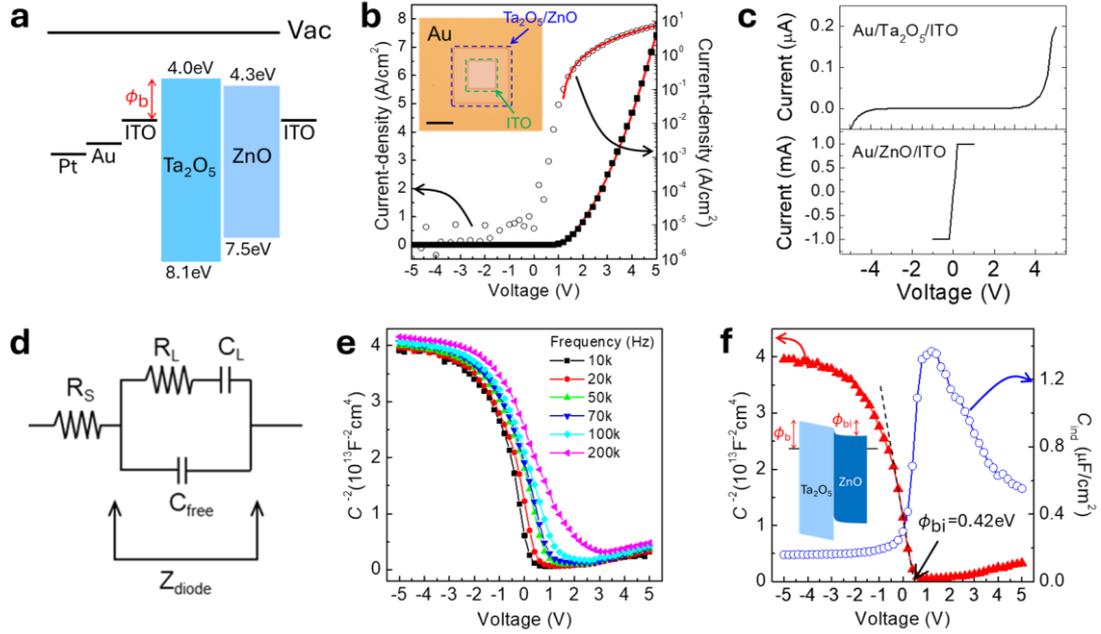

**Figure 1.** The characterization of the Ta$_2$O$_5$/ZnO hetero-oxide diode. a) Band diagram of the Ta$_2$O$_5$/ZnO Schottky diode. The $\phi_b$ indicates the Schottky energy barrier. The hetero-oxide structure is grown by in-situ sputtering at room temperature. b) Current-density versus voltage using the Schottky diode. The left and right y-axis indicate linear and log-scale current densities, respectively. The inset shows an optical microscopy image of the diode. Scale bar, 50μm. c) Current-voltage characteristics of Au/Ta$_2$O$_5$/ITO (top) and An/ZnO/ITO (bottom) devices. The current compliance was set to 1mA. d) Equivalent circuit model of the Schottky diode. $R_s$ is the series resistance, $R_L$ is the localized resistance, $C_L$ is the localized capacitance, and $C_{free}$ is the capacitance due to free charges. Z$_{diode}$ indicates the frequency-dependent impedance. e) Reciprocal of the square of capacitance with multi frequencies ranging from 10k Hz to 200k Hz. f) Extracted frequency-independent reciprocal of the square of capacitance (left-y axis) and the capacitance (right-y axis). The $\phi_{bi}$ is the built-in potential, interpolated from the linear dashed line at $C^{-2} = 0$.

where the three unknowns of $C_f$, $C_L$, and $R_L$ are calculated using the three equations for $C(\omega)$. Figure 1e plots the measured reciprocal of the square of capacitance ($1/C^2$) with multi frequencies from 10 kHz to 200 kHz. The three unknowns are determined by choosing three frequencies (e.g., 20kHz, 100kHz, and 200kHz) from the plots. Then, the frequency-independent capacitance ($C_{ind}$) is defined by the sum of $C_L$ and $C_{free}$. Figure 1f represents the $C_{ind}$, where the built-in potential of 0.42 eV is defined at the interface between Ta$_2$O$_5$ and ZnO, as illustrated by the inset.

**Details of Band Structures and High Blocking Voltage Capabilities**

Figure 2a illustrates the energy-band structures of the Schottky diode under forward and reverse biases. The conduction band offset between $Ta_2O_5$ and ZnO is about 0.3 eV. A relatively low band offset can facilitate electron injection from ZnO to $Ta_2O_5$. Due to the offset, electrons accumulate at the $Ta_2O_5$/ZnO interface under forward bias, creating a space charge region. The electrons injected into the $Ta_2O_5$ diffuse and drift towards the Au contact due to the electric field within $Ta_2O_5$ (see Note 2 in Supporting Information). Under forward bias, the voltage is mostly dropped across the $Ta_2O_5$/ZnO junction and $Ta_2O_5$. At the junction, the conduction band of the ZnO at bends downward, resulting in the accumulation of electrons. Free electrons ($n_{free}$), expressed by $n_{free} = \int_{E_{CBO}}^{\infty} n_{acc}\, dE$ (where $E_{CBO}$ is the conduction band offset at the $Ta_2O_5$/ZnO interface), are injected from ZnO to $Ta_2O_5$ and flow through $Ta_2O_5$ to the Au. This decides the amplitude of the forward current.

Figure 2b shows the reverse current density ($J_{rev}$) versus voltage up to 65V, measured over 10 cycles. The red dashed line and blue symbol represent the initial and final measurements, respectively. The curve increases slightly during the cycles. The trend is saturated, not significant enough to break down the Schottky barrier. Here the reverse current is determined by three types of charge transport: field emission ($J_{FE}$), thermionic-field emission ($J_{TFE}$), and thermionic emission ($J_{TE}$). $J_{TFE}$ and $J_{TE}$ can be suppressed when the Schottky barrier is relatively high. The room-temperature-grown $Ta_2O_5$ has defects, which induce charge transport by trap-assisted tunneling, thereby contributing to $J_{FE}$ [20]. The reverse current does not exponentially increase, indicating that the barrier thinning effect does not occur in the structure. This is because the electric field is evenly distributed across the 200-nm-thick $Ta_2O_5$ layer, mitigating the thinning effect typically observed in thin-film metal-semiconductor Schottky-junctions [7].

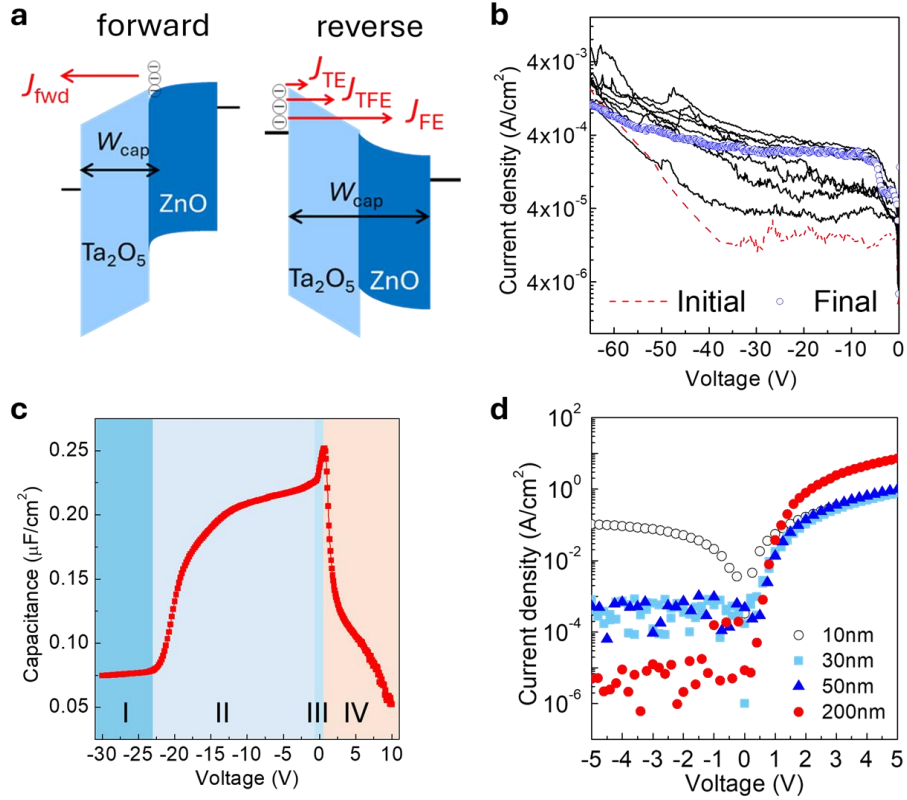

**Figure 2.** High Blocking Voltage in the $Ta_2O_5$/ZnO Structure. a) Illustrative band structures of the $Ta_2O_5$/ZnO Schottky diode under forward and reverse biases. $W_{cap}$ is the capacitance width. $J_{fwd}$ is the forward current density. $J_{TE}$, $J_{TFE}$, and $J_{FE}$ are thermionic, thermionic-field emission, and field-emission current densities, respectively. b) Reverse current-density curves during 10 continuous measurements. c) Measurement of capacitance-voltage characteristics using the Schottky diode with four regions. I: Fully depleted $Ta_2O_5$/ZnO, II: Partially depleted ZnO, III: Charge accumulation at the interface of $Ta_2O_5$/ZnO, IV: Strong charge accumulation at the interface of $Ta_2O_5$/ZnO. d) Current-density versus voltage for $Ta_2O_5$/ZnO Schottky diodes with $Ta_2O_5$ thicknesses were 10nm, 50nm, 200nm, and 500nm.

Figure 2c presents the capacitance-voltage characterization to aid understanding of the band structure via applied voltages. In the region-I, the curve is nearly flat, indicating fully depleted $Ta_2O_5$ and ZnO layers. In the region-II, the depleted region in the ZnO decreases as the reverse bias is reduced. At near-zero bias in the region-III, the ZnO is negatively charged due to electrons, which accumulate at the $Ta_2O_5$/ZnO interface. The capacitance rises to its peak value, determined by the thickness of the accumulation layer. Figure S2 (Supporting Information) shows the hysteresis measurement in the diode, indicating a small window due to charge trapping. Therefore, although the room-temperature process can introduce defect states at the interface, they can be negligible. In the region-IV, electrons accumulate at the interface and the accumulation depth extends into the ZnO layer, causing the capacitance to decrease again.

Figure 2d shows the comparison of the current-voltage characteristics of $Ta_2O_5$/ZnO Schottky diodes with different $Ta_2O_5$ thicknesses ($t_{Ta2O5}$) of 10nm, 30nm, 50nm, and 200nm. Charges, by a diffusion-drift mechanism (Figure S4 and Note 2 in Supporting Information), flow through $Ta_2O_5$. The voltage drops ($V_{drop}$) mainly occur in the $Ta_2O_5$ layer, indicating that the drift mechanism is dominant in the $Ta_2O_5$ layer. As $t_{Ta2O5}$ increases, the reverse current is significantly suppressed due to a relatively long path for charge transport. Numerically, the forward current is expected to be reduced due to the decreased electric field in $Ta_2O_5$. However, the experiment shows an improvement in the forward current. This is understood by the effect of the rise in substrate temperature during long time rf-magnetron sputtering,[21] which can lead to a reduction of defects near the $Ta_2O_5$/ZnO interface, enhancing the charge injection. To validate the analysis of the thickness-dependent trend, Figure S3 provides the statistical analysis of the Schottky diodes with the different $Ta_2O_5$ thicknesses.

**Half-wave Rectification and RF-DC Conversion**

Figure 3a shows the schematic of the half-wave rectifier circuit using the $Ta_2O_5$/ZnO Schottky diode. A plastic-circuit board was used to connect the Schottky diode to a waveform generator, a load resistor ($R_L$) of 1MΩ, and an oscilloscope (see Experimental section). Figure 3b indicates the half-wave rectification output voltage ($V_o$) under a 1 Hz sine-wave input ($V_{IN}$) with a peak-to-peak ($V_{p-p}$) of 20V. The colored box highlights the rectification of the negative bias region. Figure 3c shows the $V_o$ under a 1kHz sine-wave $V_{IN}$ with the same $V_{p-p}$, where the colored box indicates that the negative bias is not fully rectified. Figure 3d is an extended graph of the red box in Fig. 3c, showing the decay of discharging due to resistance and capacitance components in the diode structure. The capacitance is determined by $Ta_2O_5$ and electron accumulation as described in Fig. 2c. The red line represents the fitted resistive-capacitive (RC) delay curve, fitted by the exponential function of $e^{-t/\tau}$ ($\tau$: time constant of $2.86 \times 10^{-4}$). By further increasing the frequency up to 4 MHz, Figures 3e to 3g show that $V_o$ is characterized by direct current (DC) conversion. This is due to the increase in RC delay, as illustrated in Fig. 3h. The RC delay is determined by the charge distribution in the Schottky diode under forward bias. Hole charges are blocked by the $Ta_2O_5$, resulting in a positive charge at the metal interface. $X_o$ is the junction between the metal and $Ta_2O_5$. The charge $Q_1$ is formed between $X_{Ta2O5}$ and $X_{acc}$ due to electron accumulation. The charge $Q_2$ between $X_{Ta2O5}$ and $X_{Drift}$ is determined by the diffused electrons. The sum of $Q_1$ and $Q_2$ equals $Q_+$, with $Q_1$ and $Q_2$ controlled by material properties such as carrier concentration in ZnO, the $E_{CBO}$, and the density of defects in $Ta_2O_5$. The $t_{cap}$ is the thickness of the junction capacitance generated across the region between $X_{Drift}$ and $X_{acc}$. A relatively strong field forms from $X_{Drift}$ to $X_o$, accelerating electrons towards the metal. When the diode is reverse-biased, $Q_1$ and $Q_2$ are discharged due to RC delay, which characterizes the radio frequency (RF) to DC conversion observed in Figs. 3e to 3g. A high responsivity is important for RF-to-DC conversion, and Figure S5 and Note 3 (Supporting Information) indicate the estimated responsivity (A/W) in the Schottky diode, which is higher than 20.

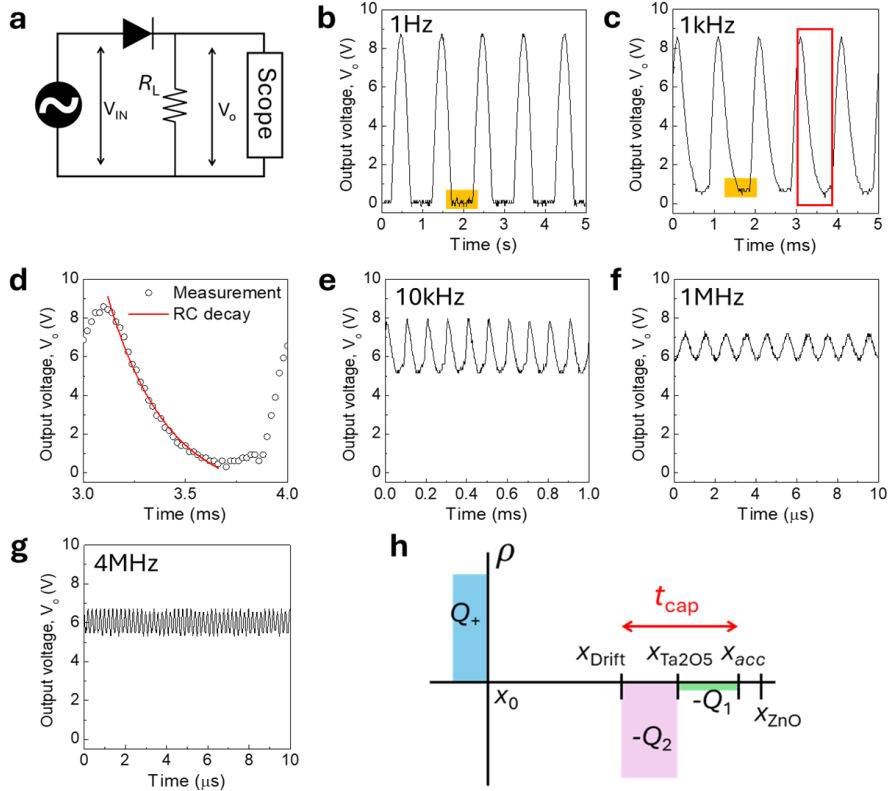

**Figure 3.** Half-wave Rectification and RF-DC Conversion. a) Schematic of the half-wave rectifier circuit using the $Ta_2O_5$/ZnO Schottky diode. $V_{IN}$: Input voltage, $R_L$: Load resistor, $V_o$: Ouput voltage. The experiment setup is described in Experiment section. b) to g) Measured $V_o$ using a sine-wave $V_{IN}$ with varying frequencies. d) shows the extended plot marked by the red box in c). The red solid line is the fitted exponential line due to resistive-capacitive discharging effect. h) Illustration of the charge distribution in the Schottky diode under forward bias. $Q_+$ represent the positive charge at $X_0$. $Q_1$ denotes the accumulated electron charges at the $Ta_2O_5$/ZnO interface. $Q_2$ is the charges in the depleted $Ta_2O_5$ region. $X_0$ is the Au/$Ta_2O_5$ junction, $X_{Drift}$ is the starting point of the drift region in $Ta_2O_5$, $X_{Ta2O5}$ is the $Ta_2O_5$/ZnO junction, and $X_{acc}$ is the endpoint of charge accumulation in ZnO.

**All-Oxide Schottky Diode and Rectifying Resistive Memory**

Figure 4a shows the rectification current-voltage curve for the all-oxide Schottky diode, illustrated by the inset. The top ITO is grounded while the applied voltage is swept at the bottom ITO. Figure S6 and Note 4 (Supporting Information) show the calculation of the Schottky barrier from the observation of reverse saturation current using the diode. Using Richardson's law, the Schottky barrier at the $ITO/Ta_2O_5$ interface is estimated to be about 0.63 eV, which is lower than 1.2 eV barrier at the $Au/Ta_2O_5$ interface. The relatively low barrier increases the reverse current due to enhanced electron injection by $J_{FE}$, $J_{TFE}$, and $J_{TE}$. It can compromise relatively high-reverse voltage operation. However, it shows reliable rectification within the reduced voltage range of -1V to 3V. In Fig. 4b, the optical bandgaps for $Ta_2O_5$ and ZnO are estimated to be 4.13 eV and 3.21 eV, respectively. Figure 4c) shows the measured transmittance using the all-oxide Schottky diode. It exhibits a transmittance of approximately 83.4% in the visible wavelength range of 400nm to 700nm. The demonstration without using pristine metals such as Au and Pt supports its application in inexpensive and transparent electronics, including wearable devices and transparent energy harvesting systems, which require wireless power control units with RF-DC converters.

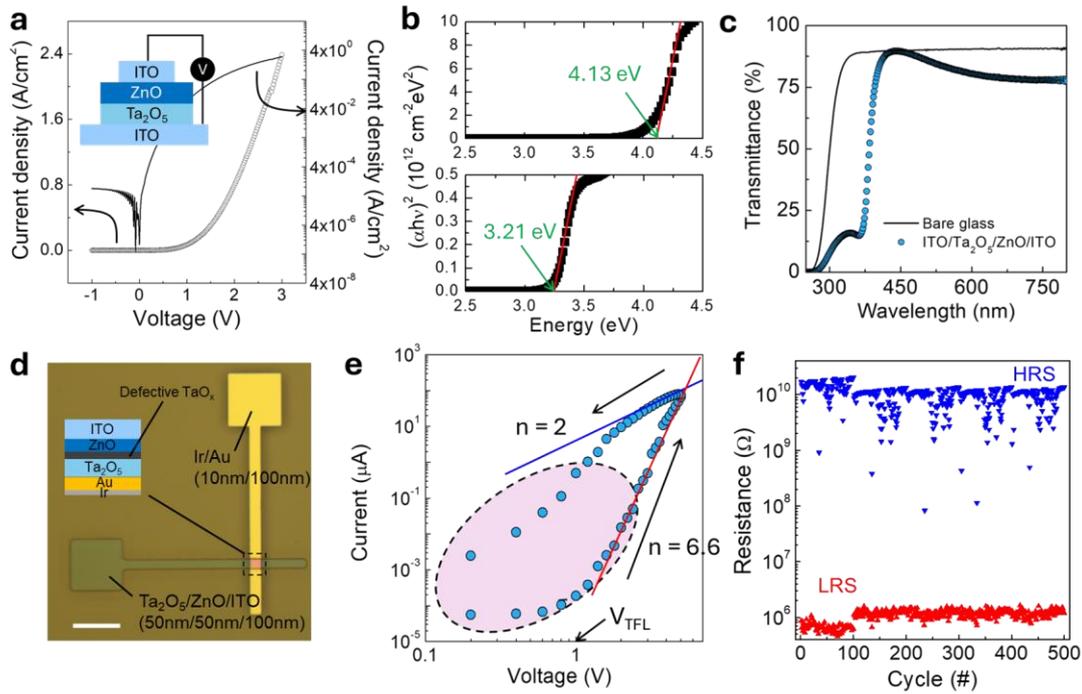

**Figure 4.** Transparent Schottky Diode and Resistive Memory Structure. a) Current density versus voltage measurement for an all oxide $Ta_2O_5$/ZnO Schottky diode. The inset shows the schematic of the diode structure. b) Optical bandgap characteristics of $Ta_2O_5$ (top) and ZnO (bottom) thin films. c) Transmittance of the transparent Schottky diode (symbols). The solid line represents the transmittance of a bare glass substrate. d) Optical microscope image of a crossbar rectifying resistive memory structure. The inset shows the thin film stack of Ir/Au/$Ta_2O_5$/$TaO_x$/ZnO/ITO. Scale bar, 50μm. e) Hysteresis characterization using the crossbar device. The dashed ellipse highlights the region associated with charge trapping and detrapping mechanisms. f) Retention test of the memory window between low-resistance state (LRS) and high-resistance state (HRS) over 500 cycles.

Figure 4d shows the optical microscope image of a crossbar Schottky diode structure fabricated on a glass substrate. The top metal is grounded while the bottom metal is swept by the applied voltage. The bottom metal consists of double layers of Ir/Au with thicknesses of 10nm/100nm, where Ir serves as the adhesive layer on the glass substrate. During the in-situ deposition of the top thin-film layers, a 5 nm defect-rich $TaO_x$ layer is sandwiched between the $Ta_2O_5$ and ZnO layers using the reactive sputtering method, where positive oxygen vacancies play a role in trapping electrons [22]. Figure S7 (Supporting Information) shows the measured transmittance of the $TaO_x$, which shows optical bandgap narrowing compared to that of $Ta_2O_5$, implying that $TaO_x$ is defect-rich near the conduction band minimum.[23] The inset illustrates the thin-film stack at the cross region, which has an area of 10µm×10µm. Using the diode, Figure 4e shows the current-voltage hysteresis measurement, demonstrating resistive memory behavior. Figure S8 (Supporting Information) illustrates the mechanism behind the formation of the resistive memory window. At a negative bias, the ionized oxygen defects in the $TaO_x$ layer remain positively charged without significant electron trapping, thereby establishing the interfacial polarization at the $TaO_x$/ZnO junction. During the voltage ramping up, the shallow defect states are filled by electrons. At a forward voltage less than the trap-filled limit voltage ($V_{TFL}$) of approximately 1V, the conduction band of the ZnO bends downward at the interface of ZnO/$TaO_x$, exposing electrons to deep trap states in the $TaO_x$. When the applied voltage exceeds $V_{TFL}$, the curve is governed by trap-limited SCLC with the relationship of $I_d \propto V^n$,[24] where n is 6.7 during the voltage ramping up. The process of trap filling suppresses the polarization in the $Ta_2O_5$, and the field across the $Ta_2O_5$ drives electrons to the Schottky metal contact. As the voltage increases, the trap states are continuously filled, and the polarization switches to the reverse direction. At a relatively high voltage, the traps in the $TaO_x$ can be fully occupied by electrons. Therefore, the $TaO_x$ layer is considered a trap-free interface. At this stage, the voltage is estimated to be higher than 5V; however, the device was not tested at voltages higher than 5V to prevent breakdown. During voltage ramp-down, the current-voltage curve ($n = 2$) behaves similar to the $Ta_2O_5$/ZnO Schottky diodes. Below $V_{TFL}$, the curve deviates from the guideline for $n = 2$ (increasing), indicating that the $TaO_x$ returns to a positively charged state due to charge detrapping. The switch of polarization at the $TaO_x$ interface generates the resistive memory characteristics.

Figure 4f represents the retention test of charge trapping and detrapping processes during hysteresis measurements across 500 consecutive cycles. Figure S9 (Supporting Information) shows the full set of the measurements across the bias range from -5V to +5V. The ratio of the high resistance state (HRS) to low resistance state (LRS) is approximately $10^4$, which is sufficiently high to reflect weight updates determined by conductance changes during the gradient descent optimization method for memory-based machine learning.[25] Figure S10 (Supporting Information) shows the optimized bias range to achieve a highest ratio of the HRS to LRS. The reliable switching characteristics suggest that this concept is a suitable candidate for realizing simple crossbar resistive memory chips, not demanding switching components such as transistors.

## Conclusion

Room-temperature, in-situ grown $Ta_2O_5$/ZnO Schottky-diodes are demonstrated, showing their high performance with an on/off ratio higher than $10^8$. A reverse blocking voltage is estimated to be at least 65V. The Schottky barrier at the metal/$Ta_2O_5$ interface accommodates a wide range of metal work functions from Pt to ITO, enabling the creation of all-oxide Schottky-diodes with over 80% transparency in the visible spectrum. Furthermore, I demonstrate the versatility of these diodes in applications such as half-wave rectifiers, RF-DC conversion, and crossbar resistive memory, highligting their current rectification capabilities. The demonstrations present a promising method for integrating the Schottky diodes into the back-end of chips, utilizing low-cost, low-temperature processes and high througput, which are crucial for reducing environment impact and promoting sustainability.

## Acknowledgements

This preprint does not include acknowledgements. Information on financial support and grants will be provided in the final published article.

## Author contributions statement

Jihun Lim was responsible for designing the experiments and conducting the sample measurements. In addition, Jihun Lim analyzed the data and prepared and wrote the manuscript.

## Data availability statement

The data and materials that support the findings of this study are available from the corresponding author upon reasonable request.

## Conflict of Interest

The authors declare no conflict of interest.